\documentclass[a4paper,11pt]{article}
\usepackage[utf8]{inputenc}

\usepackage{geometry}
\usepackage{xcolor}
\usepackage{amsmath, array, amssymb, amsfonts,amsthm}
\usepackage{xfrac}

\usepackage[french, english]{babel}
\usepackage[all]{xy}

\usepackage{lmodern}

\usepackage{sectsty}
\sectionfont{\centering}
\subsectionfont{\centering}
\subsubsectionfont{\centering}

\usepackage{booktabs}
\usepackage{caption}
\usepackage{dsfont}
\usepackage{mathtools}

\usepackage[hidelinks]{hyperref}

\usepackage{textcomp}

\usepackage{multicol}
\usepackage[square,numbers,sort,compress,semicolon,merge]{natbib}
\usepackage{notoccite}
\renewcommand\P{\mathcal{P}}
\newcommand\M{\mathcal{M}}

\newcommand\doubleR{\mathbb{R}}

\renewcommand\1{\textbf{1}}

\renewcommand\H{\mathcal{H}}

\renewcommand\L{\mathcal{L}}

\newcommand\U{\mathcal{U}}
\newcommand\SO{\mathcal{SO}}

\newcommand\K{\mathcal{K}}

\newcommand\W{\mathcal{W}}

\newcommand\rarrow{\rightarrow}

\newcommand\LieG{\mathfrak{g}}
\newcommand\LieH{\mathfrak{h}}

\newcommand\so{\mathfrak{so}}
\newcommand\co{\mathfrak{co}}

\newcommand\h{\widehat}

\newcommand\w{\wedge}
\renewcommand\d{\partial}

\renewcommand\-{^{-1}}

\renewcommand\1{\mathds{1}}

\DeclareMathOperator{\Tr}{Tr}

\theoremstyle{definition}

\hypersetup{
pdfauthor={J. François, S. Lazzarini, T. Masson},
pdftitle={Nucleon spin decomposition and differential geometry},
pdfsubject={},
pdfcreator={pdflatex},
pdfproducer={pdflatex},
pdfkeywords={}
}

\renewcommand{\thefootnote}{\fnsymbol{footnote}}

\begin{document}

\title{Weyl gravity and Cartan geometry}
\author{J. Attard,${\,}^a$ \hskip 0.7mm J. François${\,}^b$\footnote{Supported by a Riemann fellowship.} \hskip 1.5mm
and \hskip 0.7mm S. Lazzarini${\,}^a$}
\date{}

\maketitle
\begin{center}

\vskip -1cm
${}^a$ Centre de Physique Théorique,\\
Aix Marseille Université \& Université de Toulon \&  CNRS UMR 7332,\\
13288 Marseille, France\\[3mm]
${}^b$ Riemann Center for Geometry and Physics,\\
Leibniz Universität Hannover,\\
Appelstr. 2, 30167 Hannover, Germany
\end{center}

\begin{abstract}
We point out that the Cartan geometry known as the second-order conformal structure provides a natural differential geometric framework underlying gauge theories of conformal gravity. We are concerned by two theories: the first one 
will be the associated Yang-Mills-like Lagrangian, while the second, inspired by~\cite{Wheeler2014}, will be a slightly more general one which will relax the conformal Cartan geometry. The corresponding gauge symmetry is treated within the BRST language. We show that the Weyl gauge potential is a spurious degree of freedom, analogous to a Stueckelberg field, that can be eliminated through the dressing field method. We derive sets of field equations for both the studied Lagrangians. For the second one, they constrain the gauge field to be
the `normal conformal Cartan connection'. Finally, we provide in a Lagrangian framework a justification of the identification, in dimension $4$, of the Bach tensor with the Yang-Mills current of the normal conformal Cartan connection, as proved in \cite{Korz-Lewand-2003}.
\end{abstract}


\vspace{1mm}


\vspace{1.5mm}


\renewcommand{\thefootnote}{\arabic{footnote}}

\begin{multicols}{2}

\section*{Introduction}  

In $1918$-$19$, H. Weyl, trying to devise a ``truly infinitesimal geometry'' that generalizes Riemann's,\footnote{H.~Weyl did so by requiring that not only the directions of vectors at distant points a manifold couldn't be compared without a non-canonical choice of connection, as in Riemann's geometry, but also that neither could be their lengths. This he called ``scale freedom'' and then ``gauge freedom''. He thereby originated the notion of gauge symmetry, which would reveal its deepness within quantum mechanics few years later, with the posterity we know.}
came up with a spacetime manifold equipped with what today we would call a conformal class of metric: a metric defined up to positive local rescalings. The natural scale-invariant Lagrangian he proposed (of Yang-Mills type, as we could say anachronistically) intended to unify gravity and electromagnetism~\cite{Weyl1918, Weyl1919}.
The theory turned out to be incompatible with the basic experimental fact of the stability of atomic spectra. But still to this day, scale invariance retains theoretical interest, as witnessed by its importance {\em e.g.} in string theory and conformal field theory, among many other topics.

In particular, the Lagrangian for Weyl gravity 
\begin{align}
\label{Weyl _lagrangian}
\L_{\text{Weyl}} = -\Tr(W \w *W) 
=  -\tfrac{1}{2}W_{\mu\nu\rho\sigma} W^{\mu\nu\rho\sigma} dV
\end{align}
introduced by Bach in 1921~\cite{Bach1921} and constructed with the Weyl tensor $W$, is still actively investigated. 
Solutions of its field equation, the Bach equation, are under study to connect the theory to empirical data  and see if it can rival General Relativity (GR). In particular its viability as an alternative to dark matter and dark energy is still under scrutiny, as is its viability as a quantum gravity theory. See the reviews \cite{Mannheim2006, Mannheim2012} and references therein to get only a sample of the significant literature on the subject.

After the $1956$ pioneering work of Utiyama on the gauging of an arbitrary Lie group and its first treatment of gravitation as a gauge theory of the Lorentz group \cite{Utiyama1956}, in the late $1970$'s, several authors investigated the question of the gauge structure of gravity (and supergravity) \cite{Cham-West1977, Townsend1977, McDowell-Mansouri1977, West1978, Stelle-West1979}. 
 During the same period, some of them studied the gauging of the $15$-parameter conformal group extending the Poincaré group, and its supersymmetric counterpart as well \cite{Ferber-Freund1977, CrispimRomao1977, Kaku_et_al1977, Ferrara-et-al1977, Kaku-et-al1978}. For a general review see {\em e.g.}~\cite{Blagojevic:2013xpa}.

Following a more abstract differential geometric approach, authors \cite{Harnad-Pettitt1, Harnad-Pettitt2, Harnad-Pettitt3} already gave a gauge formulation of conformal gravity within the framework of higher-order frame bundles~\cite{Kobayashi}. The relevant geometry is known as the second-order conformal structure. However, it is better to use an equivalent formulation in terms of Cartan geometry~\cite{Sharpe, Kobayashi}, which allows a matrix treatment much closer to the usual gauge field framework familiar to physicists. 

As is well known, the geometry of connections on principal fiber bundles is an appropriate mathematical setting for dealing with Yang-Mills gauge theories. Because of its strong link to the spacetime manifold $\M$, Cartan geometry provides a natural framework that properly addresses the peculiarity of gravitation among the other interactions. Thus, it would perfectly fit the geometry underlying gauge theories of gravitation, in particular that of Weyl gravity.  
Accordingly, our aim is to show that the second-order conformal structure is the Cartan geometry underlying a genuine gauge formulation of conformal gravity containing Weyl gravity as a special case.

Moreover, inspection of the explicit field equations obtained in~\cite{Wheeler2014} raises the issue whether the field variables could be pieced together into a single object, namely the conformal Cartan connection. Because of the possible geometry underlying Weyl gravity, it may be relevant to give an account of this aspects.

\medskip
The paper is organized as follows. In section \ref{Second-order conformal structure}  
 we give a brief description of the second-order conformal structure. 
  In section \ref{Conformal gauge theories} we write the most natural Yang-Mills like Lagrangian, and a slightly generalized version. We show why the Weyl gauge potential of dilation can be considered as a spurious degree of freedom, and can be suppressed 
thanks to the so-called dressing field method; the latter is consistent with the locality principle. We also derive field equations and 
show that they single out the normal conformal Cartan connection as gauge field. In section \ref{Discussion} we make contact with some papers in the literature, in particular with~\cite{Wheeler2014} and in addition we will justify the equivalence between the Bach equation and the Yang-Mills current of the normal conformal Cartan connection as found in \cite{Korz-Lewand-2003}. Then we conclude.  Appendices give some details on how gauge invariance restricts the choices of Lagrangians, as well as a brief recap of the dressing field method.

\section{Second-order conformal structure}  
\label{Second-order conformal structure}  

We refer to \cite{Sharpe} and to \cite{Kobayashi, Ogiue} for a detailed mathematical presentations of Cartan geometry and higher-order frame bundles respectively. Here we just sketch the necessary material to follow our scheme.

The whole structure is modeled on the Klein pair of Lie groups $(G, H)$  where $G= O(2,m) / \lbrace \pm I_{m+2} \rbrace$ 
and $H$ is the isotropy group 
such that the corresponding homegeneous space is the compactified Minkowski space $(S^{m-1}\times S^1)/\mathbb{Z}^2 \simeq G/H$.
The group $H$ has the following factorized matrix presentation
\begin{align}
\label{struct_group}
 H = K_0\, K_1=\left\{\! \begin{pmatrix} z &  0 & 0  \\  0  & S & 0 \\ 0 & 0 & z^{-1}  \end{pmatrix}\!\!  \begin{pmatrix} 1 & r & \frac{1}{2}rr^t \\ 0 & \1 & r^t \\  0 & 0 & 1\end{pmatrix} \!\right\}
\end{align} 
where $z\in \mathsf{W}=\doubleR^*_+$, $S\in SO(1, m-1)$ and $r\in \doubleR^{m*}$.
Here ${}^t$ stands for the $\eta$-transposition, namely for the row vector $r$ one has $r^t = (r \eta^{-1})^T$ (the operation ${}^T\,$ being the usual matrix transposition), and $\doubleR^{m*}$ is the dual of $\doubleR^m$. 
We refer to $\mathsf{W}$ as the Weyl group of rescaling. Obviously $K_0\simeq CO(1, m-1)$, and $K_1$ is the abelian group of inversions (or conformal boosts). 

Infinitesimally we have the Klein pair  $(\LieG, \LieH)$ of graded Lie algebras~\cite{Kobayashi}. They decompose respectively as, $\LieG=\LieG_{-1}\oplus\LieG_0\oplus\LieG_1 \simeq \doubleR^m\oplus\co\oplus\doubleR^{m*}$, and $\LieH=\LieG_0\oplus\LieG_1 \simeq \co\oplus\doubleR^{m*}$. In matrix notation,
\begin{multline*}
\mathfrak{g}=\left\{ 
\begin{pmatrix} \epsilon &  \iota & 0  \\  \tau  & v & \iota^t \\ 0 & \tau^t & -\epsilon  \end{pmatrix}   
\right\} 
\supset
\LieH = \left\{ \begin{pmatrix} \epsilon &  \iota & 0  \\  0  & v & \iota^t \\ 0 & 0 & -\epsilon  \end{pmatrix}\right\}
\end{multline*} 
with $ (v-\epsilon\1)\in \mathfrak{co},\ \tau\in\mathbb{R}^m,\ \iota\in\mathbb{R}^{m*}$ and the $\eta$-transposition of the  column vector $\tau$ is $\tau^t = (\eta\tau)^T$.
The graded structure of the Lie algebras, $[\LieG_i, \LieG_j] \subseteq \LieG_{i+j}$, $i,j=0,\pm 1$ with the abelian Lie subalgebras $[\LieG_{-1}, \LieG_{-1}] = 0 = [\LieG_1, \LieG_1]$, is automatically handled by the matrix commutator.
\medskip

The second-order conformal structure is a Cartan geometry $(\P, \varpi)$ where $\P=\P(\M, H)$ is a principal bundle over $\M$ with structure group $H=K_0 K_1$, and $\varpi\in \Omega^1(\U, \LieG)$ is a  (local) Cartan connection $1$-form on $\U \subset \M$. The curvature of $\varpi$ is given by the structure equation, $\Omega=d\varpi +\tfrac{1}{2}[\varpi, \varpi]=d\varpi +\varpi^2 \in \Omega^2(\U, \LieG)$ (the wedge product is tacit: $\varpi^2=\varpi\w\varpi$). Both have matrix representations
\begin{align}
\label{Cartan_connect}
\varpi=\begin{pmatrix} a & \alpha & 0 \\ \theta & A & \alpha^t \\ 0 & \theta^t & -a   \end{pmatrix} \text{ and } 
\Omega=\begin{pmatrix} f & \Pi & 0  \\ \Theta & F & \Pi^t \\ 0 & \Theta^t & -f \end{pmatrix}.
\end{align}

One can single out the so-called \emph{normal} conformal Cartan connection (which is unique) by imposing the constraints 
\begin{align}
\label{normality}
\Theta=0\ \text{  (torsion free)}  \quad \text{and} \quad  {F^a}_{bad}=0.
\end{align}
 Together with the $\LieG_{-1}$-sector of the Bianchi identity $d\Omega+[\varpi, \Omega]=0$, \eqref{normality} implies $f=0$ (trace free), so that the curvature of the normal Cartan connection reduces to 
 \vspace{-2mm}
 \begin{align*}
 \Omega=\begin{pmatrix} 0 & \Pi & 0  \\ 0 & F & \Pi^t \\[0.5mm] 0 & 0 & 0 \end{pmatrix}.
 \end{align*}
 From the normality condition ${F^a}_{bad}=0$ in \eqref{normality}, follows that $\alpha$ 
 has components (in the $\theta$ basis of $\Omega^\bullet (\U)$)
 \begin{align}
 \label{Schouten}
 {\alpha}_{ab}=-\frac{1}{(m-2)} \left( R_{ab} - \frac{R}{2(m-1)}\eta_{ab} \right)
 \end{align}
 where $R$ and $R_{ab}$ are the Ricci scalar and Ricci tensor associated with the $2$-form $R=dA+A^2$. 
   In turn, from \eqref{Schouten}  follows that  
   \begin{align*}
   F:= R + \theta \alpha + \alpha^t\theta^t=W
   \end{align*}
    is the Weyl $2$-form. By the way, in the gauge $a=0$, $\Pi:=d\alpha +\alpha A=D\alpha$ looks like what we can call the Cotton $2$-form. 
    
\medskip

The principal bundle $\P(\M, H)$ is a second order $G$-structure, a reduction of the second order frame bundle $L^2\M$; it is thus a ``2-stage bundle''. The bundle $\P(\M,H)$ over $\M$ can also be seen as a principal bundle $\P_1:=\P(\P_0,K_1)$ with structure group $K_1$ over $\P_0:=\P(\M,K_0)$.

\section{Conformal gauge theories}  
\label{Conformal gauge theories}  

\subsubsection*{Yang-Mills conformal Lagrangian } 

In this geometrical setting given by the above principal bundle $\P(\M, H)$, consider the Cartan connection~\eqref{Cartan_connect} $\varpi$ as gauge field and its curvature $\Omega$ as field strength. A physical theory for treating the dynamics of the gauge field is given by a choice of gauge invariant action functional. For instance, one may take a $\H$-gauge invariant Lagrangian $m$-form, with $m=$ dim$\M$ and $\H :=\left\{ \gamma: \U \subset \M\rarrow H \right\}$ the gauge group.

The most obvious and natural choice is to write the Yang-Mills prototype Lagrangian
\begin{align}
\label{YM-conf}
\L_{\text {YM}}(\varpi)&= \Tr(\Omega \w *\Omega), \\
	     &= \Tr(F\w *F) + 4\Pi\w *\Theta + 2f \w *f.\notag
\end{align}
At this stage some care is required. Indeed when $\H$ acts, so too does the Weyl gauge group of rescalings, $\W:=\{ z: \U \subset \M \rarrow  \textsf{W} \}$. In particular, its action on $\varpi$ implies  $\theta^W = z \theta$. Hence, given a $p$-form $B$, the Hodge operator transforms under $\W$ according to
\begin{align*}
(*B)^W = z^{m-2p} *B.
\end{align*}
Therefore, the $\H$-invariance of the Lagrangian~\eqref{YM-conf} requires to restrict oneself to a spacetime $\M$ of dimension $m=4$; \footnote{ This peculiarity of dimension $m=4$ is very similar to the requirement of the conformal invariance of the Maxwell Lagrangian density $\L_{\text{Maxwell}}(F,g)= F *_g F$. Indeed, $\L_{\text{Maxwell}}(F,z^2 g)= z^{m-4}\L_{\text{Maxwell}}(F,g)$ implies $\L_{\text{Maxwell}}(F,z^2 g) = \L_{\text{Maxwell}}(F,g)$ for all $z\in\W$ if $m=4$.}
this will be assumed throughtout the rest of the paper.

\smallskip
Along the line suggested by~\cite{Wheeler2014}, one can also choose the slightly more general Lagrangian, which relaxes the conformal Cartan geometry
\begin{align}
\label{Gen}
\MoveEqLeft{\L_{\text{gen}}(\varpi)=} \notag\\
& c_1 \Tr(F\w *F) + c_3 \Pi\w *\Theta + c_2  f \w *f
\end{align}
with $c_1$, $c_2$ and $c_3$ arbitrary constants.
 
Some remarks are in order. First, the discrepency from the case $4c_1=c_3=2c_2$ is not quite natural with respect to the underlying geometry. Second, let $\delta_0$ and $\delta_1$ be the infinitesimal actions of the gauge subgroups $\K_0$ and $\K_1$ respectively. One has (see appendix \ref{Symmetries of the Lagrangians})  $\delta_0\L_{\text{gen}}=0$ since each of the three terms in \eqref{Gen} is separately $\K_0$-invariant, but 
\vspace{-1mm}
\begin{align}
\label{Gen1}
\MoveEqLeft{\delta_1 \L_{\text{gen}}=} \\
& (4c_1-c_3) \Tr\big(\Theta \kappa \w *F\big)
			 + (c_3- 2c_2) \kappa \Theta \w *f \notag 								\end{align}
where $\kappa$ is the infinitesimal $\K_1$ parameter (i.e an infinitesimal conformal boost).  This vanishes only if $\Theta=0$, or if $4c_1=c_3=2c_2$.   The latter case is of course $\L_{\text{gen}}=c_1 \L_{{\text {YM}}}$, the natural choice dictated by the conformal geometry.
Confronted with this problem one can adopt three strategies.
\medskip

First,  one could restore full $\H$-invariance by restricting to a torsion free geometry $\Theta=0$ from the very beginning. This reduces the Lagangian \eqref{Gen} to 
\begin{align}
\label{Wheeler}
\L_\text{W}(\varpi) :=  c_1 \Tr(F\w *F) + c_2\  f \w *f.
\end{align}

As well, if one is willing to allow for torsion, one could \emph{state} that the $\K_1$ gauge group of conformal boost doesn't act, thus breaking by hand the gauge symmetry from $\H$ to $\K_0$. 

Finally, the third route consists in erasing the $\K_1$ gauge symmetry by means of the so-called $K_1$-valued \emph{dressing field} $u_1$, as described in \cite{FLM2015_II} (see appendix \ref{The dressing field method}). 
 This amounts to a local reduction of $\P(\M, H)$ to the subbundle $\P(\M, K_0)$. The dressing of $\varpi$ and $\Omega$ respectively gives
\begin{align}
\label{u_1 comp fields}
\varpi_1 &:= u_1\- \varpi u_1 + u_1\-du_1 = \begin{pmatrix} 0 & \alpha_1 & \!\!0 \\ \theta & A_1 &\!\!\alpha_1^t \\ 0 & \theta^t &\!\!0   \end{pmatrix}, \notag \\[-4mm]
\\
\Omega_1 &:= u_1\-\Omega u_1 = d\varpi_1\!+\varpi_1^2=\begin{pmatrix} f_1 & \Pi_1 & \!\!0  \\ \Theta & F_1 & \!\!\Pi_1^t \\ 0 &\Theta^t &\!\!-f_1 \end{pmatrix}. \notag
\end{align}
These are \emph{not} gauge transformations (see \ref{The dressing field method} and \cite{Fournel-et-al2014, FLM2015_I, FLM2015_II}) but $\K_1$-invariant composite fields. Nevertheless, they still transform as $\K_0$-gauge fields.  Thus, in $\varpi_1$, the $1$-form $A_1$ is the genuine spin connection. 

In the normal case, that is imposing the condition \eqref{normality}, $\alpha_1$ is  the Schouten $1$-form with components given, \emph{mutadis mutandis}, by \eqref{Schouten}. Since by dressing the gauge invariance of $a_1=0$ is guaranteed, the entry $\Pi_1=d\alpha_1+ A_1\alpha_1$ is the Cotton $2$-form. A further consequence is that $F_1$ is the Weyl $2$-form. 
\medskip

By the way, given that $\L_{\text {YM}}(\varpi^{\gamma_1})=\L_{\text {YM}}(\varpi)$, for $\gamma_1:\U\rarrow K_1 \  \in \K_1$. And using the formal resemblance between gauge transformation and dressing, one has $\L_{\text {YM}}(\varpi)=\L_{\text {YM}, 1}(\varpi_1)$ with
\begin{align}
\label{YM-conf1}
\L_{\text {YM}, 1}(\varpi_1)&= \Tr(\Omega_1 \w *\Omega_1) \\
	     &= \Tr(F_1\w *F_1) + 4\Pi_1\w *\Theta + 2f_1 \w *f_1. \notag
\end{align}
This Lagrangian is $\K_1$-invariant because it is constructed with $\K_1$-invariant fields, the only true residual gauge symmetry being $\K_0$ (Lorentz $\times$ Weyl). Furthermore, it gives a field equation for the gauge field $\varpi_1$ which unfolds as three equations only:  respectively for the vielbein field $\theta$, the spin connection $A_1$ and $\alpha_1$. The Weyl gauge potential of dilation, $a$ in the previous writing of the theory, was a spurious degree of freedom, compensated by an `artificial' $\K_1$ gauge symmetry.\footnote{ The dressing field method is shown to be here the inverse of the Stueckelberg procedure, which aims at implementing a gauge symmetry by adding the so-called Stueckelberg field. In the  situation at hand, $a$ is such a Stueckelberg field indeed. See appendix in \cite{FLM2015_I} for a discussion.}

\medskip
The analogue of \eqref{Gen} for the dressed variables,
\begin{align}
\label{Gen_1}
\MoveEqLeft{\L_{\text{gen},1}=}\\
& c_1 \Tr(F_1\w *F_1) + c_3\ \Pi_1\w *&\Theta 
							 + &c_2\  f_1 \w *f_1 \notag
\end{align}
 is invariant under the Lorentz gauge group $\SO \subset \K_0$, but not under the Weyl gauge group  $\W$ (see appendix \ref{The dressing field method}). Indeed, if $\delta_W$ is the infinitesimal Weyl action with parameter $\epsilon \in$ Lie$\W$ ($z=\exp(\epsilon)$), then
\begin{align*}
\delta_W \L_{\text{gen}, 1}= (4c&_1-c_3) \Tr\big(\Theta(\d\epsilon\!\cdot\!e\-) \w *F_1\big) \notag\\[1mm]
 +& (c_3 - 2c_2)
(\d\epsilon\!\cdot\!e\-)\Theta \w *f_1.
\end{align*}
This vanishes only if $\Theta=0$, or if $4c_1=c_3=2c_2$, that is $\L_{\text{gen},1}=c_1 \L_{\text{YM},1}$, the natural choice for which $\delta_W\L_{\text{YM}, 1}=0$ as expected. 
  
But now we have no choice, we cannot freeze the action of the Weyl gauge group $\W$, neither by decree nor by dressing. In order to preserve the $\W$-invariance, one \emph{must} require $\Theta=0$, the torsionless condition. Implementing the latter in \eqref{Gen_1} restricts oneself to
\begin{align}
\label{Wheeler_1}
\L_{\text{W},1}(\varpi_1)=  c_1 \Tr(F_1\w *F_1) + c_2\  f_1 \w *f_1
\end{align}
as a theory for the gauge potential and field strength
\begin{align*}
\varpi_1=\begin{pmatrix} 0 & \alpha_1 & 0 \\ \theta & A_1 & \alpha_1^t \\ 0 & \theta^t & 0   \end{pmatrix}, \quad \Omega_1=\begin{pmatrix} f_1 & \Pi_1 & 0  \\ 0 & F_1 & \Pi_1^t \\ 0 & 0 & -f_1 \end{pmatrix}. 
\end{align*}

\subsubsection*{Normality and field equations} 

The field equations deriving from $\L_{\text{YM}, 1}$ \eqref{YM-conf1} are obtained by varying the corresponding action with respect to (w.r.t.) the dressed Cartan connection $\varpi_1$ (see \eqref{u_1 comp fields}). Two contributions must be considered: one is the standard Yang-Mills term, the other comes from variation of the Hodge-$*$ operator, defined w.r.t. the coframe basis $\{\theta\}$ for differential forms: 
\begin{align*}
\delta_{\varpi_1} S_{\text{YM}, 1} = \int \left( \Tr(\delta \varpi_1 \w D_1*\Omega_1) + \delta \theta \w T^{\Omega_1} \right) =0
\end{align*}
where $D_1:= d  + [\varpi_1,\  ]$ and  $ T^{\Omega_1}$  is the energy-momentum $3$-form of $\Omega_1$. 
Thanks to the non-degeneracy of the Killing form and taking into account the various sectors of the Lie algebra, one gets three equations w.r.t. the respective three gauge fields
\begin{align*}
\delta \alpha_1&: & & D*\Theta  - *F_1 \w \theta + \theta \w *f_1 = 0, \\
\delta A_1&: & & D*F_1 - * \Theta \w \alpha_1 + \alpha_1^t \w *\Theta^t  \\
& & & \hskip 1.5cm + \theta \w *\Pi_1 - *\Pi_1^t \w \theta =0, \\
\delta \theta&: & &  D*\Pi_1 - *f_1 \w \alpha_1+ \alpha_1 \w *F_1 = -\tfrac{1}{2} T^{\Omega_1},
\end{align*}
where $D:= d + [A_1, \ ]$. 

\medskip
The field equations for $\L_{\text{W}, 1}$  \eqref{Wheeler_1} are a special case of  those of $\L_{\text{gen}, 1}$ \eqref{Gen_1} (see appendix \ref{General field equations}). They read 
\begin{align*}
\delta \alpha_1&: &  2c_1 *F_1 \w \theta - c_2 \theta \w *f_1 &= 0,  \\
\delta A_1&: &  D*F_1  &=0, \\
\delta \theta&: & c_2\,  *f_1 \w \alpha_1 - 2c_1 \alpha_1 \w *F_1  &=  c_1 T^{F_1} + c_2T^{f_1}.
\end{align*}
Dropping out the subscript ``$1$'' for convenience, one has in components, 
\begin{align}
 2c_1\ {F^c}_{b, ca} - c_2\  f_{ab}&=0, \label{eq_alpha} \\
 D^c {F^d}_{a, cb} &=0, \label{eq_A}\\
 c_2\ \alpha_{a, c}{f_b}^c - 2c_1\ \alpha_{c,d} {{F^c}_{a, b}}^d&= c_1 T_{ab}^{F_1} + c_2T_{ab}^{f_1} \label{eq_theta}
\end{align}
with the two energy-momentum tensors,
\begin{align*}
 T_{ab}^{F_1}&=  \tfrac{1}{4} {F^i}_{j, cd}{{F^j}_{i,}}^{cd} \eta_{ab} +  {F^i}_{j, bc}{{F^j}_{i,}}^{cd} \eta_{da} , \\[2mm]
  T_{ab}^{f_1}&=  \tfrac{1}{4} f_{cd}f^{cd} \eta_{ab} +  f_{bc}f^{cd} \eta_{da} .
  \end{align*}
A remarkable fact is that the field equations~\eqref{eq_alpha} select the  (dressed) normal conformal Cartan connection as gauge field, \emph{provided} that $c_2\neq 2c_1$. Let us prove this. 

From the Bianchi identity $D\Omega_1=[\Omega_1, \varpi_1]$ which is easily written in matrix form, the $\LieG_{-1}$-sector reads $d\Theta=(F_1-f_1\1)\theta - A_1\Theta$. Since $\Theta=0$ this reduces to
$(F_1-f_1\1)\theta=0,$ or in components ${F^a}_{[b, cd]}=f_{[cd}{\delta^a}_{b]}.$ 
By contracting over $a$ and $b$ and remembering that ${F^a}_{a, cd}=0$ since $F\in \so(1, 3)$, one has
\begin{align*}
{F^a}_{c, ad}-{F^a}_{d, ac}=-2 f_{cd}.
\end{align*}
Now the antisymmetric part of \eqref{eq_alpha} is 
\begin{align*}
c_1({F^c}_{a, cb}-{F^c}_{b, ca}) + c_2 f_{ab}=0.
\end{align*}
Combining these two equations, we end up with
\begin{align*}
(c_2-2c_1)f_{ab}=0.
\end{align*}
Now the point in writing the linear combination \eqref{Gen_1}, thus \eqref{Wheeler_1}, was to depart from the natural (and rigid) geometric case $c_2=2c_1$. So the above equation implies $f_{ab}=0$, which in turn implies that \eqref{eq_alpha} reduces to 
\begin{align}
\label{normality_1}
{F^c}_{acb} =0. 
\end{align}
In other words, the field equations of $\L_{W, 1}$ single out the dressed normal Cartan connection as gauge field. 

Since in this case $\alpha_1$ is the Schouten $1$-form (a function of $A_1$ through solving \eqref{normality_1}), it is not an independent field variable. 
Furthermore, since $A_1 \in \so(1, 3)$ and $\Theta=0$, the spin connection $A_1$ is a function of the vielbein field $e={e^a}_\mu$. Thus, the only  independent gauge field in $\varpi_1$ is the  vielbein $1$-form $\theta=e\cdot dx ={e^a}_\mu dx^\mu $.

It is quite easy to see that it induces a conformal class of metrics $\{ g \}$. Indeed from \eqref{BRST_1} in appendix \ref{The dressing field method}, one has that the gauge BRST variation of $\varpi_1$ provides
\begin{align*}
s_L \theta=-v_L \theta, \quad \text{and} \quad s_W \theta = \epsilon \theta,
\end{align*}
where $v_L \in \so(1, 3)$ is the Lorentz ghost, and $\epsilon$ is the Weyl ghost. So that defining a metric by $g:=e^T \eta e$, one has the infinitesimal gauge transformations,
\begin{align*}
s_Lg &= (s_Le)^T \eta e +e^T\eta s_L e = -e^T(v_L^T\eta + \eta v_L)e=0 , \\
s_Wg &= (s_We)^T \eta e +e^T\eta s_W e = 2\epsilon (e^t\eta e) = 2\epsilon g.
\end{align*}
In other words, at the finite level, one has $g^{\gamma_0}=z^2g$. This means that the true degrees of freedom of the theory described by $\L_{\text{W}, 1}$ \eqref{Wheeler_1} are those of a conformal class of metric $\{g \}$ ($\tfrac{m(m+1)}{2} - 1=9$ in dimension $m=4$). 
\medskip

Moreover, in dimension $m=4$, the tensor $T^{F_1}_{ab}$ vanishes identically, see \cite{Lovelock1967, Lovelock1970}. It is then easily seen that while \eqref{eq_alpha}  enforces the normality, combining \eqref{eq_A} and \eqref{eq_theta} provides particular solutions of the Bach equation, 
\begin{align}
\label{Bach1}
2D_d  D^c {F^d}_{a, bc}+  \alpha_{c,d} {{F^c}_{a, b}}^d =0
\end{align}
but do not exhaust them. 

\section{Discussion} 
\label{Discussion} 

Aiming at finding the vacuum Einstein equations from conformal gravity, the author of \cite{Wheeler2014} (see also \cite{Trujillo2013}) \emph{starts} with the Lagrangian $\L_\text{W}$ \eqref{Wheeler}, that is setting $c_3=0$ in \eqref{Gen}. With this choice of Lagrangian he needs to \emph{assume}, first that $\K_1$ does not act (breaking of the gauge symmetry by hand), and second that $\Theta=0$ for the field equations to enforce normality. Subsequently, he also requires the gauge fixing condition  $a=0$ (there referred to as the `Riemann gauge') for the Cartan connection $\varpi$.

Obtaining the Lagrangian $\L_{\text{W}, 1}$ \eqref{Wheeler_1} by redefining the fields through the dressing field method has several advantages. Indeed, the vanishing of the (dressed) Weyl potential $a_1$ and the $\K_1$-invariance are simultaneously guaranteed by the dressing construction. Furthermore, $\L_{\text{W}, 1}$ is $\SO$-invariant, and requiring the invariance under $\W$ \emph{imposes} $\Theta=0$ right away.
Then, the field equations for $\L_{\text{W}, 1}$, directly select the normal conformal Cartan connection as gauge field.  

Suppose that the  choice of the constants in $\L_{\text{W}, 1}$ is taken to be the natural one with respect to the underlying geometry of the second-order conformal structure, $c_2=2c_1$. Then the field equations fail to select the (dressed) normal conformal Cartan connection. 

\bigskip  
The authors of \cite{Korz-Lewand-2003}  made the mathematical observation that, in dimension $4$, the Bach tensor can be identified with the Yang-Mills current of the normal conformal Cartan connection in what they refer to as the `natural gauge', that is with $a=0$ (in our notation). This observation receives a clear meaning in the dressing field scheme and in a Lagrangian field theory approach.
 
Indeed, starting with the normal subgeometry of the second-order conformal structure $\P(\M, H=K_0 K_1)$, and after dressing (w.r.t. the $K_1$ direction), the normal conformal Cartan connection associated to $\P(\M, K_0)$ and its curvature read
\begin{align*}
\varpi_1=\begin{pmatrix} 0 & \alpha_1 & 0 \\ \theta & A_1 & \alpha_1^t \\ 0 & \theta^t & 0   \end{pmatrix}, \quad \Omega_1=\begin{pmatrix} 0 & \Pi_1 & 0  \\ 0 & F_1 & \Pi_1^t \\ 0 & 0 & 0 \end{pmatrix},
\end{align*}
with $\alpha_1$ the Schouten $1$-form, $A_1$ the spin connection, $\Pi_1=D\alpha_1$ the Cotton $2$-form and $F_1$ the Weyl $2$-form. The natural Yang-Mills Lagrangian then reduces to
\begin{align}
\label{Lagrangian_YM_normal}
\L_{\text{YM}, 1}(\varpi_1)= \Tr(\Omega_1 \w *\Omega_1) = \Tr (F_1 \w *F_1).
\end{align}
Varying of the  action w.r.t. $\varpi_1$ gives
\begin{align*}
\delta_{\varpi_1} S_{\text{YM}, 1}= \int \Tr(\delta \varpi_1 \w D_1*\Omega_1) + \delta \theta \w T^{\Omega_1}=0,
\end{align*}
where the energy-momentum  $T^{\Omega_1}$ reduces to  $T^{F_1}$, which vanishes identically ($m=4$). 
Then, the field equation is just the Yang-Mills equation 
\begin{align*}
D_1*\Omega_1=0,
\end{align*}
the Yang-Mills current of \cite{Korz-Lewand-2003}.
  Unfolding it we get,
\begin{align*}
\delta \alpha_1&: \quad   *F_1 \w \theta = 0,  \\
\delta A_1&: \quad D*F_1  + \theta \w *\Pi_1 - *\Pi_1^t \w \theta^t= 0, \\
\delta \theta&: \quad   D*\Pi_1 + \alpha_1 \w *F_1  = 0.
\end{align*}
After dualizing through the Hodge $*$ and dropping out once more the subscript $1$ for convenience, one has
\begin{align*}
\delta \alpha_1&: \quad   {F^c}_{a, cb}=0,  \\
\delta A_1&: \quad D^j {F^a}_{b, rj} + \Pi_{b, rj}\eta^{aj}+ \eta^{aj}\Pi_{j, br}= 0, \\
\delta \theta&: \quad   D^c\Pi_{a, bc} + \alpha_{cd} {{F^c}_{a, b}}^d= 0.
\end{align*}
The first equation above is identically satisfied because it gives back one of the two conditions of normality assumed from the very beginning. Using   the $\LieG_0$-sector of the Bianchi identity $D_1\Omega_1=0$, which is the well-known result $D_d {F^d}_{a, bc} + \Pi_{a, bc}=0$, one shows that the second equation above is also identically satisfied. 
Thus, the only equation giving information is that stemming from the variation of the tetrad field, 
\begin{align}
\label{Bach}
D^cD_{[b}\alpha_{c],a} + \alpha_{cd} {{F^c}_{a, b}}^d= 0.
\end{align}
This is nothing but the Bach equation (in an alternative form equivalent to \eqref{Bach1} in dimension $4$). 

In other words, in dimension 4, the field equation for $\L_{\text{YM}, 1}$ \eqref{Lagrangian_YM_normal} is the Yang-Mills equation,
\begin{align} 
\label{equiv}
D_1*\Omega_1 = 
\begin{pmatrix} 0 & D*D\alpha_1 + \alpha_1\w *F_1 & 0 \\0 &0 & * \\ 0 & 0 &0  \end{pmatrix} =0 
\end{align}
and is equivalent to the Bach equation \eqref{Bach}.

This was naturally expected since 
$\L_{\text{YM}, 1}$ \eqref{Lagrangian_YM_normal} is nothing but the Lagrangian $\L_{\text{Weyl}}$ \eqref{Weyl _lagrangian} of Weyl gravity, and as noted above, the vielbein $\theta$ is the only independent field in the dressed normal conformal Cartan connection $\varpi_1$. Thus, variation of $\L_{\text{YM}, 1}$ under $\varpi_1$ giving $D_1*\Omega_1=0$ is the same as variation of $\L_{\text{Weyl}}$ under $\theta$ giving the Bach equation as usual.

\section*{Conclusion}  
\label{Conclusion}  

 In this paper we highlighted the second-order conformal structure as the global geometrical framework underlying gauge conformal theories of gravity, and the conformal Cartan connection as the natural gauge potential. 
 
We have shown that the Weyl potential $a$ for dilation is a Stueckelberg-like field whose spurious degrees of freedom can be absorbed through the dressing field method. 
This provides an advantageous substitute to the gauge fixing a=0 imposed in \cite{Wheeler2014}, and results in the effective local reduction of the second-order conformal structure to the first-order conformal structure.

We have discussed two choices of Lagrangians, a Yang-Mills type Lagrangian dictated by the conformal geometry and a more generalized one, inspired by \cite{Wheeler2014}, which relaxes the conformal geometry. In the latter case, we have stressed that the field equations select the unique (dressed) normal conformal Cartan connection as gauge  potential. 

Furthermore, in this geometrical setup, we have provided a Yang-Mills  theory which justifies (see Lagrangian \eqref{Lagrangian_YM_normal} and eq.\eqref{equiv}) the identification, in dimension $4$, \cite[see there section 3]{Korz-Lewand-2003} of the Bach tensor with the Yang-Mills current of the normal conformal Cartan connection. 

\appendix 

\section{Symmetries of the Lagrangians} 
\label{Symmetries of the Lagrangians}

 Under the gauge group $\H :=\left\{ \gamma: \U \subset \M\rarrow H \right\}$, the curvature $\Omega$ transforms by the adjoint: $\Omega^\gamma=\gamma\-\Omega \gamma$. This is why the choice $\L_\text{YM}(\varpi)=\Tr(\Omega\w*\Omega)$ as    $\H$-invariant  Lagrangian is natural. To consider other possibilities, it is interesting to pay attention to the action of the subgroups of $\H$. 

Consider the  gauge transformations 
\begin{align*}
\gamma_0:\U \rarrow K_0 \quad \text{and} \quad \gamma_1:\U \rarrow K_1
\end{align*}
elements of the subgroup $\K_0$  and $\K_1$ respectively.  Given the matrix representation \eqref{struct_group}, one has 
\begin{align*}
\Omega^{\gamma_0}&=\begin{psmallmatrix} f & z\-\Pi S & 0  \\ S\-\Theta z &  S\- F S & S\- \Pi^t z\- \\ 0 &  z\Theta^t S & -f  \end{psmallmatrix}
\quad \text{ and} \\[2mm]
 \Omega^{\gamma_1}&=\begin{psmallmatrix} f-r\Theta & \Pi - r(F-f\1) - r\Theta r + \tfrac{1}{2}rr^t\Theta^t  & 0\\  \Theta & \Theta r + F - r^t\Theta^t & * \\ 0 &\Theta^t & * \end{psmallmatrix}. \notag
\end{align*}

By inspection one sees that each term in the natural Lagrangian \eqref{YM-conf}
\begin{align*}
\L_\text{YM}(\varpi)= \Tr(F\w *F) + 4\Pi\w *\Theta + 2f \w *f
\end{align*}
 are separetely $\K_0$-invariant. 
This means that even the more general Lagrangian
\begin{align}
\label{gen}
\MoveEqLeft{\L_\text{gen}= } \\
& c_1 \Tr(F\w *F) + c_3\ \Pi\w *\Theta + c_2\  f \w *f \notag
\end{align}
with $c_1$, $c_2$ and $c_3$ arbitrary constants, is $\K_0$-invariant. 
Thus, so is the Lagrangian \eqref{Wheeler} considered in \cite{Wheeler2014, Trujillo2013}. 
\medskip

The $\K_1$-invariance imposes more restrictions.
For simplicity, consider an infinitesimal conformal boost $r=\kappa$ (an inversion). The linear variation of $\Omega$ is
\footnote{Along with the linear variation of the Cartan connection $\varpi$, they can be both obtained by writing the $\K_1$ sector of the BRST algebra of the theory (the subscript $i$ stands for inversion) 
\vspace{-1mm}
\begin {align*}
s_i \varpi = -dv_i - [\varpi, v_i], \quad s_i\Omega=[\Omega, v_i]\quad \text{ and } \quad s_i v_i = -\tfrac{1}{2}[v_i, v_i]=-v_ i^2=0,\quad \text{ with } \quad v_i=\begin{pmatrix} 0 & \kappa & 0 \\ 0 & 0& \kappa^t \\ 0 & 0 & 0 \end{pmatrix},
\end{align*}
where $v_i$ is the anticommuting ghost field associated with infinitesimal conformal boosts. See \cite{FLM2015_II} for an extensive treatment of the BRST algebras associated with the second-order conformal structure $\P(\M, H)$. 
}
\begin{align*}
\delta_1 \Omega = \begin{pmatrix} -\kappa \Theta & -\kappa(F-f\1) & 0  \\ 0 & \Theta \kappa - \kappa ^t \Theta^t & * \\ 0 & 0& *  \end{pmatrix}.
\end{align*}
It is then easy to show that 
\begin{align*}
\delta_1 \L_\text{YM} = 4 \Tr\left(\Theta \kappa \w *F\right) - 4 \kappa F\w*\Theta = 0,
\end{align*}
as expected. But the general Lagrangian transforms as
\begin{align}
\label{gen1}
\delta_1 \L_\text{gen} = (4c_1 - c_3)  &\Tr \big(\Theta\, \kappa \w *F\big)\notag\\ 
			 & + (c_3 - 2c_2) \kappa\, \Theta \w *f.  							
\end{align}
This vanishes only if $\Theta=0$, or if $4c_1 = c_3 = 2c_2$.\footnote{These relations can also be found by requiring the nilpotency of the BRST operator, $s_i^2\L_\text{gen}=0$.}
  The latter case is $\L_\text{gen}=c_1 \L_\text{YM}$, the natural choice dictated by the geometry.

 If one doesn't want to be restricted to a torsion free geometry, and nevertheless wants to restore  full gauge-invariance, then the so-called dressing field method is the way forward. See  \cite{Fournel-et-al2014, FLM2015_I, FLM2015_II} for details, and the following for a brief recap.

\section{The dressing field method} 
\label{The dressing field method}

The gauge group of a gauge field theory  is defined as
$\H:=\left\{ \gamma :\U \rarrow H\right\}$ and acts on itself by $\gamma_1^{\gamma_2}=\gamma_2\- \gamma_1 \gamma_2$ for any $\gamma_1, \gamma_2  \in \H$. It  acts on the gauge potential and the field strength  according to,
\begin{align}
\label{GT}
A^\gamma=\gamma\-A\gamma + \gamma\*d\gamma, \qquad F^\gamma=\gamma\-F\gamma. 
\end{align}

Suppose the theory also contains a (Lie) group-valued field
$u:\U\rarrow G'$ defined by its transformation under  $\H'=\left\{
  \gamma':\U\rarrow H'\right\}$, where $H' \subseteq H$ is a subgroup, given by $u^{\gamma'}:={\gamma'}\-u$.
One  can then define the following {\em composite fields}, 
\begin{align}
\label{comp_fields}
\h A := u\- A u + u\-du,  \qquad \h F :=u\-F u.
\end{align}
The Cartan structure equation holds for the dressed curvature $\h F=d\h A + {\h A}{\,}^2$. 

Despite the formal similarity with \eqref{GT}, the composite fields \eqref{comp_fields} are not mere gauge transformations since $u\notin \H$, as witnessed by its transformation property under $\H'$ and the fact that in general $G'$ can be different from $H$. This  implies that the composite field $\h A$ does no longer belong to the space of local connections. 

As is easily checked, the composite fields \eqref{comp_fields} are
$\H'$-invariant and are only subject to residual gauge transformation
laws in $\H \setminus \H'$. 
In the case $H'=H$, these composite fields are $\H$-gauge invariants.
\medskip

It is easy to show that the BRST gauge algebra pertaining to a pure gauge theory is modified by the dressing as
\vspace{-1mm}
\begin{align}
\label{dressed_BRST}
&s\h A =-\h D \h v=-d\h v - [\h A, \h v\,], \quad   s\h F=[\h F, \h v\,], \notag\\[1mm]
&\text{and} \quad  s\h v =-\tfrac{1}{2}[\,\h v ,\h v\,]= - \h v{\,}^2,
\end{align}
upon defining the {\em composite ghost}
\begin{align}
\label{dressed_ghost}
\h v:= u\- v u + u\- s u \ .
\end{align}
It encodes the infinitesimal residual gauge symmetry, if any. If $\h v=0$,  the BRST algebra \eqref{dressed_BRST} becomes trivial, thus expressing the gauge invariance of the composite fields. 
\bigskip

As for the case of the second-order conformal structure, the gauge group is $\H=\K_0\K_1$, and it is possible to reduce $\H$ down to $\K_0$ by dressing in the $\K_1$-direction. Consider the field $u_1:\U \rarrow K_1$~with
\begin{align*}
u_1 = \begin{pmatrix} 1 & q & \tfrac{qq^t}{2} \\ 0  & \1 & q^t \\ 0 & 0 & 1  \end{pmatrix}.
\end{align*}
Imposing on the Cartan connection $\varpi$ the gauge-like condition $\chi(\varpi^{u_1})=a^{u_1}=a-q\theta=0$ and solving for $q$, one can check that
$u_1^{\gamma_1}=\gamma_1\-u_1$ for $\gamma_1\in \K_1$. 
Then $u_1$ is indeed a $\K_1$-dressing field which can be used to form the $\K_1$-invariant composite fields \footnote{In order to stick to \cite{FLM2015_II} the $\h{\ }\,$ has been dropped out as in the main text.}
\begin{align*}
\varpi_1:=u_1\- \varpi u_1 + u_1\-du_1, \quad \text{and} \quad \Omega_1:=u_1\-\Omega u_1
\end{align*}
whose matrix forms are displayed in \eqref{u_1 comp fields}. These fields are well behaved as $\K_0$-gauge fields, so that the dressing amount to a (local) reduction of the second-order conformal structure $\P(\M, H)$ to the first-order conformal structure $\P(\M, K_0)$. See \cite{FLM2015_II} for details.

Furtermore, one can also check that not only $\chi\big((\varpi^{\gamma_1})^{u_1^{\gamma_1}}\big)=\chi(\varpi^{u_1})$, which is the gauge-like condition's $\K_1$-invariance that enforces the dressing transformation law for $u_1$, but also that $\chi\big((\varpi^{\gamma})^{u_1^\gamma}\big)=\chi(\varpi^{u_1})$ for $\gamma \in \H$. Which means that the condition $a_1:=a^{u_1}=0$ in the dressed field $\varpi_1$ displayed in \eqref{u_1 comp fields}, is fully $\H$-invariant. 
\medskip

The BRST algebra of $\P(\M, H)$ is modified. The initial full ghost is, 
\begin{align*}
v=v_0+v_i=v_W + v_L+v_i= \begin{psmallmatrix} \epsilon & \kappa & 0 \\[1mm] 0 & v_L & \kappa^t \\[1mm] 0 & 0& -\epsilon  \end{psmallmatrix}  \  \in \text{Lie}\H
\end{align*}
with $v_0=v_W+v_L \in$ Lie$\K_0$ the decomposition in the Weyl and Lorentz sector, and $v_i \in$ Lie$\K_1$ the ghost of conformal boost. 

\noindent
After dressing the composite ghost is
\begin{align}
\label{v_1}
v_1 := u_1\- v u_1 + u_1\- s u_1 = \begin{psmallmatrix}  \epsilon\ & \d\epsilon \cdot e\-  & 0 \\[1mm] 0 & v_L & (\d\epsilon \cdot e\-)^t \\[1mm] 0 & 0& -\epsilon  \end{psmallmatrix} 
\end{align}
where $\d\epsilon\!\cdot e\-= \d_\mu \epsilon\,{e^\mu}_a$ replaces the ghost of conformal boost $\kappa$. The associated modified BRST algebra~is
\begin{align}
\label{BRST_1}
s_1\varpi_1 &=-D_1v_1, & s_1v_1&=-v_1^2 \\
 s_1\Omega_1 &=[\Omega_1, v_1] & & \notag
 \end{align} 
with $s_1^2=0$. Now, since the composite ghost \eqref{v_1} admits the decomposition, 
\begin{align*}
v_1 = v_L+v'_W = \begin{psmallmatrix}  0 & 0  & 0 \\[1mm] 0 & v_L &0 \\[1mm] 0 & 0& 0  \end{psmallmatrix} + \begin{psmallmatrix}  \epsilon\ & \d\epsilon \cdot e\-  & 0 \\[1mm] 0 & 0 & (\d\epsilon \cdot e\-)^t \\[1mm] 0 & 0& -\epsilon  \end{psmallmatrix}.
\end{align*}
The algebra \eqref{BRST_1} splits into two subalgebras, 
\begin{align*}
s_L\varpi_1 &=-D_1v_L, & s_1v_L &=-v_L^2,   \\[1mm]
s_W\varpi_1 &=-D_1v'_W,  & s_1v_1 &=-{v'_W}^2 \\[1mm]
(s_L\Omega_1 &=[\Omega_1, v_L], & s_W\Omega_1 &=[\Omega_1, v'_W])
\end{align*}
with $s_L^2=0$ and $s_W^2=0$.\footnote{And $s_Ls_W+s_Ws_L=0$  since $s_Lv_W=-v_Lv_W$ and $s_Wv_L=-v_Wv_L$.}
\bigskip

In the Lorentz sector let us write explicitly, 
\begin{align*}
s_L\Omega_1= \begin{psmallmatrix}  0 & \Pi_1 v_L & 0 \\[1mm] -v_L\Theta \ \  & [F_1, v_1] \ &\  -v_L\Pi_1^t \\[1mm]  0 & \Theta^t v_L & 0 \end{psmallmatrix}.
\end{align*}
This readily gives $s_L \L_{\text{YM}, 1}=0$ since each piece in \eqref{YM-conf1} is inert under $s_L$. This also means that the more general Lagrangian
\begin{align}
\label{gen_1}
\MoveEqLeft{\L_{\text{gen},1} =} \\
& c_1 \Tr(F_1\w *F_1) + c_3\ \Pi_1\w *\Theta 
+ c_2\  f_1 \w *f_1 \notag
\end{align}
 enjoys  Lorentz invariance, $s_L\L_{\text{gen},1}=0$. 
\medskip

In the Weyl subalgebra, let us write explicitly 
\begin{align*}
s_W\Omega_1= \begin{psmallmatrix}  -(\d\epsilon \cdot e\-) \Theta\ \  & -\epsilon \Pi_1 - (\d\epsilon \cdot e\-)\big(F_1 -f_1\1 \big)\ \ & 0 \\[1mm]
\Theta \epsilon & \Theta(\d\epsilon \cdot e\-) - (\d\epsilon \cdot e\-)^t\Theta^t & * \\[2mm]
0 & \epsilon \Theta^t & *\end{psmallmatrix}.
\end{align*}
One can easily show that 
\begin{align*}
s_W \L_{\text{gen}, 1}= (4c&_1-c_3) \Tr\big(\Theta(\d\epsilon\!\cdot\!e\-) \w *F_1\big) \notag\\[1mm]
 &+ (c_3 - 2c_2)
(\d\epsilon\!\cdot\!e\-)\Theta \w *f_1
\end{align*}
which is the analogue of \eqref{Gen1} but where the infinitesimal conformal boost $\kappa$ has been replaced by $\d\epsilon \cdot e\-$.
This vanishes only if $\Theta=0$, or if $4c_1=c_3=2c_2$.
\footnote{These relations are also found by requiring the nilpotency of the BRST operator, $s_W^2\L_{\text{gen},1}=0$.}
  The latter case is $\L_{\text{gen},1}=c_1 \L_{\text{YM},1}$, the natural choice for which $s_W\L_{\text{YM}, 1}=0$ is expected. 

  \section{General field equations} 
\label{General field equations}
  
For the sake of completeness, we here provide the field equations for $\L_{\text{gen}, 1}$ stemming from the variations $\delta \alpha_1$, $\delta A_1$ and $\delta \theta$ respectively,
\begin{align*}
 c_3 D*\Theta- 4c_1 *F_1 \w \theta + 2c_2 \theta \w *f_1 =& 0,  \\
 2c_2 D*F_1 -\tfrac{c_3}{2} \big(    *\Theta \w \alpha_1 - \alpha_1^t \w*\Theta^t\big)& \\
+ \tfrac{c_3}{2} \big( \theta \w *\Pi_1 - *\Pi_1\w \theta^t \big)=&0, \\
 c_3 D*\Pi_1 -2 c_2  *f_1 \w \alpha_1+ 4c_1 \alpha_1 \w *F_1  =& -2 T^{EM}
\end{align*}
where $D:=d+[A_1, \ ]$. Applying the Hodge star operator to get equations for $1$-forms, and dropping the subscript ``$1$'' for convenience, one has in components, 
\begin{align*}
c_3\ D^c{\Theta^d}_{,ac}\eta_{db} - 4c_1\ {F^c}_{b, ca} + 2c_2\  f_{ab}&=0, \\[1mm]
2c_1\ D^c {F^d}_{a, bc} + \tfrac{c_3}{2}\big( \Pi_{a, bc} \eta^{cd}  -\eta^{dc}\Pi_{c, ba}  \big) \quad & \\
-\tfrac{c_3}{2}\big(  {\Theta^d}_{, bc} \alpha_{a, e} \eta^{ec} -   \eta^{dc}\alpha_{c, e} {\Theta^n}_{bm}\eta_{na} \eta^{em} \big)
&=0, \\[1mm]
2c_2\ \alpha_{a, c}{f_b}^c  - c_3\ D^c \Pi_{a, bc} - 4c_1\ \alpha_{c,d} {{F^c}_{a, b}}^d &= 2T_{ab}^{EM}
\end{align*}
with the symmetric energy-momentum tensor,
\begin{align*}
T_{ab}^{EM}=& \  c_1\big( \tfrac{1}{4} {F^i}_{j, cd}{{F^j}_{i,}}^{cd} \eta_{ab} +  {F^i}_{j, bc}{{F^j}_{i,}}^{cd} \eta_{da} \big) \\[1mm]
 &+  c_3\big( \tfrac{1}{4} \Pi_{j, cd}\Theta^{j,cd} \eta_{ab} +  \Pi_{j, bc}\Theta^{j,cd} \eta_{da} \big) \\[1mm]
 &+  c_2\big( \tfrac{1}{4} f_{cd}f^{cd} \eta_{ab} +  f_{bc}f^{cd} \eta_{da} \big).
  \end{align*}
Notice that the last term (which is similar to the energy-momentum tensor of Electromagnetism), exists even if the gauge field of Weyl dilation $a_1$ vanishes. 

Obviously, with the natural values $4c_1=c_3=2c_2$ the above equations reduce to those of $\L_{\text{YM}, 1}$. For $c_3=0$ they provide the equations for $\L_{\text{W}, 1}$.


\begin{thebibliography}{31}
\providecommand{\natexlab}[1]{#1}
\providecommand{\url}[1]{\texttt{#1}}
\expandafter\ifx\csname urlstyle\endcsname\relax
  \providecommand{\doi}[1]{doi: #1}\else
  \providecommand{\doi}{doi: \begingroup \urlstyle{rm}\Url}\fi

\bibitem[Wheeler(2014)]{Wheeler2014}
J.T. Wheeler.
\newblock {Weyl gravity as general relativity}.
\newblock \emph{Phys. Rev.}, D90\penalty0 (2):\penalty0 025027, 2014.

\bibitem[Korzy{\'n}ski and Lewandowski(2003)]{Korz-Lewand-2003}
M.~Korzy{\'n}ski and J.~Lewandowski.
\newblock {The normal conformal Cartan connection and the Bach tensor}.
\newblock \emph{Class. Quantum Grav.}, 20\penalty0 (16):\penalty0 3745, 2003.

\bibitem[Weyl(1918)]{Weyl1918}
H.~Weyl.
\newblock {Gravitation and electricity}.
\newblock \emph{Sitzungsber. Preuss. Akad. Wiss. Berlin (Math. Phys.)},
  1918:\penalty0 465, 1918.

\bibitem[Weyl(1919)]{Weyl1919}
H.~Weyl.
\newblock {A New Extension of Relativity Theory}.
\newblock \emph{Annalen Phys.}, 59:\penalty0 101--133, 1919.
\newblock [Annalen Phys.364,101(1919)].

\bibitem[Bach(1921)]{Bach1921}
R.~Bach.
\newblock {Zur Weylschen Relativit{\"a}tstheorie und der Weylschen Erweiterung
  des Kr{\"u}mmungstensorbegriffs}.
\newblock \emph{Mathematische Zeitschrift}, 9\penalty0 (1-2):\penalty0
  110--135, 1921.
\newblock URL \url{http://dx.doi.org/10.1007/BF01378338}.

\bibitem[Mannheim(2006)]{Mannheim2006}
P.D. Mannheim.
\newblock {Alternatives to dark matter and dark energy}.
\newblock \emph{Progress in Particle and Nuclear Physics}, 56\penalty0
  (2):\penalty0 340--445, 2006.

\bibitem[Mannheim(2012)]{Mannheim2012}
P.D. Mannheim.
\newblock {Making the Case for Conformal Gravity}.
\newblock \emph{Foundations of Physics}, 42\penalty0 (3):\penalty0 388--420,
  2012.

\bibitem[Utiyama(1956)]{Utiyama1956}
R.~Utiyama.
\newblock {Invariant Theoretical Interpretation of Interaction}.
\newblock \emph{Phys. Rev.}, 101:\penalty0 1597--1607, Mar 1956.

\bibitem[Chamseddine and West(1977)]{Cham-West1977}
A.H. Chamseddine and P.C. West.
\newblock {Supergravity as a Gauge theory of supersymmetry}.
\newblock \emph{Nuclear Physics B}, 129:\penalty0 39--44, 1977.

\bibitem[Townsend(1977)]{Townsend1977}
P.~K. Townsend.
\newblock {Cosmological constant in Supergravity}.
\newblock \emph{Physical Review D}, 15:\penalty0 2802--2804, 1977.

\bibitem[MacDowell and Mansouri(1977)]{McDowell-Mansouri1977}
S.~W. MacDowell and F.~Mansouri.
\newblock {Unified Geometric Theory of Gravity and Supergravity}.
\newblock \emph{Physical Review Letters}, 38:\penalty0 739--742, 1977.

\bibitem[West(1978)]{West1978}
P~C. West.
\newblock {A geometric gravity lagrangian}.
\newblock \emph{Physics Letters}, 76B:\penalty0 569--570, 1978.

\bibitem[Stelle and West(1979)]{Stelle-West1979}
K~S. Stelle and P~C. West.
\newblock {de Sitter gauge invariance and the geometry of the Einstein-Cartan
  theory}.
\newblock \emph{J. Phys. A}, 12:\penalty0 205--210, 1979.

\bibitem[Ferber and Freund(1977)]{Ferber-Freund1977}
A.~Ferber and P.G.O. Freund.
\newblock {Superconformal Supergravity and Internal Symmetry}.
\newblock \emph{Nucl. Phys.}, B122:\penalty0 170, 1977.

\bibitem[{Crispim Romao} et~al.(1977){Crispim Romao}, A.Ferber, and
  Freund]{CrispimRomao1977}
J.~{Crispim Romao}, A.Ferber, and P.G.O. Freund.
\newblock {Unified Superconformal Gauge Theories}.
\newblock \emph{Nucl. Phys.}, B126:\penalty0 429, 1977.

\bibitem[Kaku et~al.(1977)Kaku, Townsend, and Nieuwenhuizen]{Kaku_et_al1977}
M.~Kaku, P.~K. Townsend, and P.~Van Nieuwenhuizen.
\newblock {Gauge theory of the conformal and superconformal group}.
\newblock \emph{Physics Letters}, 69B:\penalty0 304--308, 1977.

\bibitem[Ferrara et~al.(1977)Ferrara, Kaku, Townsend, and van
  Nieuwenhuizen]{Ferrara-et-al1977}
S.~Ferrara, M.~Kaku, P.~K. Townsend, and P.~van Nieuwenhuizen.
\newblock {Gauging the Graded Conformal Group with Unitary Internal
  Symmetries}.
\newblock \emph{Nucl. Phys.}, B129:\penalty0 125, 1977.

\bibitem[Kaku et~al.(1978)Kaku, Townsend, and van
  Nieuwenhuizen]{Kaku-et-al1978}
M.~Kaku, P.~K. Townsend, and P.~van Nieuwenhuizen.
\newblock {Properties of Conformal Supergravity}.
\newblock \emph{Phys. Rev.}, D17:\penalty0 3179, 1978.

\bibitem[Blagojevi{\'c} and Hehl(2013)]{Blagojevic:2013xpa}
M.~Blagojevi{\'c} and F.W. Hehl, editors.
\newblock \emph{{Gauge Theories of Gravitation. A Reader with commentaries}}.
\newblock {Imperial College Press}. World Scientific, 2013.

\bibitem[Harnad and Pettitt(1976)]{Harnad-Pettitt1}
J.P. Harnad and R.B. Pettitt.
\newblock {Gauge theories for space-time symmetries}.
\newblock \emph{J. Math. Phys.}, 17:\penalty0 1827--1837, 1976.

\bibitem[Harnad and Pettitt(1977)]{Harnad-Pettitt2}
J.P. Harnad and R.B Pettitt.
\newblock {Gauge Theory of the Conformal Group}.
\newblock In T.~Sharp and B.~Kolman, editors, \emph{{Group Theoretical Methods
  in Physics (Proceedings of the Fifth International Colloquium, Montr{\'e}al
  1976)}}, pages 277--301. Academic Press, Inc., 1977.

\bibitem[Harnad and Pettitt(1978)]{Harnad-Pettitt3}
J.P. Harnad and R.B. Pettitt.
\newblock {Gauge theories for space-time symmetries II: second order conformal
  structures}.
\newblock CRM-745, February 1978.

\bibitem[Kobayashi(1972)]{Kobayashi}
S.~Kobayashi.
\newblock \emph{{Transformation Groups in Differential Geometry}}.
\newblock Springer, 1972.

\bibitem[Sharpe(1996)]{Sharpe}
R.~W. Sharpe.
\newblock \emph{{Differential Geometry: Cartan's Generalization of Klein's
  Erlangen Program}}, volume 166.
\newblock Springer, 1996.

\bibitem[Ogiue(1967)]{Ogiue}
K.~Ogiue.
\newblock {Theory of Conformal Connections}.
\newblock \emph{Kodai Math. Sem. Rep.}, 19:\penalty0 193--224, 1967.

\bibitem[Fran\c{c}ois et~al.(2015{\natexlab{a}})Fran\c{c}ois, Lazzarini, and
  Masson]{FLM2015_II}
J.~Fran\c{c}ois, S.~Lazzarini, and T.~Masson.
\newblock {Residual Weyl symmetry out of conformal geometry and its BRST
  structure}.
\newblock \emph{JHEP}, 09:\penalty0 195, 2015{\natexlab{a}}.
\newblock doi:10.1007/JHEP09(2015)195,.

\bibitem[Fournel et~al.(2014)Fournel, Fran\c{c}ois, Lazzarini, and
  Masson]{Fournel-et-al2014}
C.~Fournel, J.~Fran\c{c}ois, S.~Lazzarini, and T.~Masson.
\newblock {Gauge invariant composite fields out of connections, with examples}.
\newblock \emph{Int. J. Geom. Methods Mod. Phys.}, 11\penalty0 (1):\penalty0
  1450016, 2014.

\bibitem[Fran\c{c}ois et~al.(2015{\natexlab{b}})Fran\c{c}ois, Lazzarini, and
  Masson]{FLM2015_I}
J.~Fran\c{c}ois, S.~Lazzarini, and T.~Masson.
\newblock {Nucleon spin decomposition and differential geometry}.
\newblock \emph{Phys. Rev.}, D91:\penalty0 045014, 2015{\natexlab{b}}.

\bibitem[Lovelock(1967)]{Lovelock1967}
D.~Lovelock.
\newblock {The Lanczos identity und its generalizations.}
\newblock \emph{{Atti Accad. Naz. Lincei, VIII. Ser., Rend., Cl. Sci. Fis. Mat.
  Nat.}}, 42:\penalty0 187--194, 1967.
\newblock ISSN 0392-7881.

\bibitem[Lovelock(1970)]{Lovelock1970}
D.~Lovelock.
\newblock {Dimensionally dependent identities}.
\newblock \emph{Mathematical Proceedings of the Cambridge Philosophical
  Society}, 68:\penalty0 345--350, 9 1970.
\newblock ISSN 1469-8064.
\newblock \doi{10.1017/S0305004100046144}.

\bibitem[Trujillo(2013)]{Trujillo2013}
J.T. Trujillo.
\newblock \emph{{Weyl Gravity as a Gauge Theory}}.
\newblock PhD thesis, Utah State U., 2013.

\end{thebibliography}
\end{multicols}

\end{document}